\newif\ifpreprint					

\ifpreprint						
 \documentstyle[prl,aps,epsfig,preprint,12pt]{revtex}
\else							
 \documentstyle[prl,aps,epsfig,twocolumn]{revtex}	
\fi							

\newcommand{\ds}{\displaystyle}
\newcommand{\dd}{\partial}

\newcommand{\be}{\begin{equation}}
\newcommand{\ee}{\end{equation}}

\newcommand{\G}{\Gamma}
\newcommand{\w}{\omega}
\newcommand{\dw}{\Delta\omega}
\newcommand{\dk}{\Delta k}
\newcommand{\tdw}{\Delta\tilde\omega}
\newcommand{\tdk}{\Delta \tilde k}
\newcommand{\g}{\gamma}

\newcommand{\W}{\Omega}

\newlength{\textwidthm}
\begin{document}
\draft

\ifpreprint						
 \preprint{Freezing Light
}
\else							
 \wideabs{						
\fi							

\title{
Freezing light via hot atoms
%
}
\author{
	Olga Kocharovskaya,$^{1,2}$
	Yuri~Rostovtsev,$^{1}$
	and Marlan~O.~Scully$^{1,3}$
}
\address{
	$^1$ 	Department of Physics and Institute for Quantum Studies, 
                          Texas A\&M University,\\
		College Station, Texas~~77843-4242\\
$^2$ Institute of Applied Physics, RAS, Nizhny Novgorod, Russia 
\\	$^3$ Max-Planck-Institut f\"{u}r Quantenoptik,	D-85748 Garching, Germany 
}
\date{\today}
%
%
\maketitle

\begin{abstract}

     We prove that it is possible to freeze a light pulse (i.e.,
to bring it to a full stop) or even to make its group velocity
negative in a coherently driven Doppler broadened atomic medium via
electromagnetically induced transparency (EIT). This
remarkable phenomenon  of the ultra-slow EIT polariton 
is based on the spatial dispersion of the refraction index 
$n(\w,k)$, i.e., its wavenumber dependence, which is due to atomic
motion and provides a negative contribution to the group velocity.  
This is related to, but qualitatively different from, 
the recently observed light slowing caused by large
temporal (frequency) dispersion.

\end{abstract}

\ifpreprint						
\bigskip\bigskip					
\fi							

\ifpreprint						
 \newpage
\else							
 } 
 \narrowtext						
\fi							


Slow group velocity in coherently driven media \cite{Harris-theor-exp}
has been shown to provide new regimes of 
nonlinear interaction with 
highly increased efficiency even for 
very weak light fields, 
high precision spectroscopy and magnetometry 
\cite{Harris-NLO-weak}. 
It has been demonstrated 
\cite{Harris-theor-exp,hau99nature,kash99prl} that EIT 
is accompanied by large frequency dispersion,
$|\w\dd n/\dd\w|\gg 1$, 
and can slow the group velocity down to 10 - 10$^2$  m/s.

     In this paper we show that, using spatial dispersion 
due to atomic motion, it is possible to freeze the light, 
$v_g=0$, or even to make its group 
velocity opposite to the wavevector, $v_g<0$ (see Eq.~(\ref{vg1})). 
We consider two different types of atomic media: 
(i) atomic beam or uniformly moving sample, 
and (ii) hot gas in a stationary cell.

Freezing of light in a stationary cell via a hot gas is especially 
intriguing (Fig.~1). The idea is to tune the driving field to
resonance with the velocity group of atoms that moves in the
direction opposite to the light pulse with velocity equal to
the light group velocity that would be supported by this group
of atoms if they were at rest.

The main result of the present paper is contained in Fig.~2 which shows that 
$v_g$ can be zero, for a pulse in a hot gas, when the drive detuning $\dw_d$ 
is properly chosen. 

As is well-known, in a medium possessing both temporal and
spatial dispersion of the refraction index, 
$n(\w,k)=1+2\pi\chi(\w,k)$, 
the group velocity of light contains two contributions,
\be
v_g\equiv Re\; \ds{d\w\over dk} = Re{ c-\w \ds{\dd n(\w,k)\over \dd k}
\over
n(\w,k)+\w\ds{\dd n(\w,k)\over \dd\w}}=\tilde v_g-v_s.   
\label{vg1}
\ee
Eq.~(\ref{vg1}) is an immediate result of differentiating the 
dispersion equation $kc=\w n(\w,k)$, i.e., $c=v_g(n+\w\dd n/\dd\w) + \w\dd n/\dd k$.
The meaning of Eq.~(\ref{vg1}) becomes clear if one turns to the equation for a field 
amplitude 
$$
\left(
c{\dd\over\dd z} +{\dd\over\dd t}
\right){\cal E} = 2\pi i\w\int dz'dt'\chi(t-t',z-z'){\cal E}(t',z'). 
$$
Using the convolution theorem to write the RHS as 
$\int d\bar kd\bar\w \chi(\bar\w,\bar k){\cal E}(\bar\w,\bar k)
\exp i(\bar\w t - \bar kz)$,
expanding the susceptibility to the first order in $\bar k$, $\bar\w$, 
noting that $\bar k$ and $\bar\w$ 
under the integral may be written in terms of $\dd/\dd z$ and $\dd/\dd t$ acting on 
${\cal E}(t,z)$ and rearranging terms we have
$$
\left(c-{2\pi\w}{\dd\chi\over\dd k}\right)
{\dd {\cal E}\over\dd z} +
\left(1+{2\pi\w}{\dd\chi\over\dd\w}\right){\dd {\cal E}\over\dd t}
 = {2\pi i\w}\chi(\w,k){\cal E} 
$$
which implies the field equation with $v_g$ given by Eq.~(\ref{vg1}),
$$
\left(
v_g{\dd\over\dd z} +{\dd\over\dd t}
\right){\cal E} =
{2\pi i k} \chi(\w,k)\left[1+\ds{2\pi\w}{\dd\chi\over\dd \w}\right]^{-1}{\cal E}. 
$$
The first term in Eq.~(\ref{vg1}), 
$\tilde v_g = Re{[c/(n+\w\ds{\dd n/ \dd\w}})]$, 
is due to frequency dispersion, and was discussed in recent 
papers \cite{Harris-theor-exp,Harris-NLO-weak,hau99nature,kash99prl}.
The second term, 
$v_s = Re{[(\w\ds{\dd n/ \dd k})/(n+\w\ds{\dd n/ \dd\w}})]$,
is due to
the effect of spatial dispersion, i.e., nonlocal response of the medium
to a probe field. 
 We study dilute systems 
 where the susceptibility is small, $|\chi(\w,k)| \ll 1$, 
 but  $v_g\ll c$, 
 as it is for all the EIT experiments carried out so far.
 As usual, we consider real-valued group velocities under the
 condition that imaginary part of $d\w/ dk$ is negligible. Otherwise
 group velocity looses its simple kinematic meaning and strong
 absorption governs or prevents 
 propagation of the light pulse through the medium.
 The latter is the reason 
 why the resonant interaction of light with a two-level medium 
 never results in an ultra-slow polariton. 

%
%
%
{\bf A mono-velocity atomic beam} or uniformly moving sample
corresponds to the simple case of spatial dispersion, so-called
drift dispersion. 
In the co-moving frame atoms are at rest, 
there is no spatial dispersion and the group 
velocity is given by the first term of Eq.~(1) alone,
$\tilde v_g$.
     The Galilean transformation to the laboratory frame,
$
k=\tilde k,\;\;\; \w=\tilde \w- \tilde k v,                                           
$ 
where $v$ is the atomic velocity, 
yields the group velocity 
$
v_g = Re (d\w/dk)= \tilde v_g  - v.  
$

Eq.~(1) yields the same result,
since
the susceptibility \cite{3,scullybook} depends only on the combination $\w+k v$, 
\be
\chi_v(\w,k)=\chi(\w+k v)= {i\mu_{ab}^2 N\over \hbar}{
n_{ab}\G_{cb} + \ds{\W^2 n_{ca}/\G^*_{ac}}
\over\G_{ab}\G_{cb}+\W^2}. 
\ee
Here 
$n_{ab} = \rho_{aa} - \rho_{bb}$, 
$n_{ca} = \rho_{cc} - \rho_{aa}$, 
$\rho_{ii}$ is the population of the $i$th level, 
$\g$ and $\g_{cb}$ are the relaxation rates of excited state 
and $c-b$ coherence respectively ($\g\gg\g_{cb}$);
$\w_{ab}$ and $\w_{cb}$ are the frequencies of the optical  
and low frequency transitions ($\w_{ab} \gg \w_{cb}$);
$\w_d$, $k_d$ and $\w$, $k$ 
are the frequency and wavenumber of the driving and probe 
fields respectively, 
$N$ is the atomic density,
$\W=|\mu_{ac} E_d|/2\hbar$ is the Rabi frequency of drive field
$(1/2)E_d\exp(i\w_d t -ik_dz)+ c.c.$, 
$\mu_{ac}$ and $\mu_{ab}$ are the dipole moments 
of $a-c$ and $a-b$ transitions respectively,
$\G_{ac} = \g + i(\dw_d + k_d v)$, 
$\G_{ab} = \g + i(\dw   + k   v)$, 
$\G_{cb} = \g_{cb} + i(\dw -\dw_d + \dk v)$,
$\dw_d = \w_d - \w_{ac}$, 
$\dw = \w - \w_{ab}$, 
$k_d = {\w_d/c}$,
$k = k_d + \w_{cb}/c + \dk$. 
We use a standard model with incoherent pump 
and loss rates ($r_c=r_b=\g_{cb}/2$), assuming 
time of flight broadening of $b-c$ transition \cite{scullybook}
(Fig.~1), so that in the absence of fields $\rho_{cc} = \rho_{bb} = 1/2$. 
According to Eqs.~(1), (2), we again obtain $v_g=\tilde v_g -v$.
The physical reason for this drifting of the pulse is that the field is basically ``seized'' 
by the atoms in the form of atomic polarization. 

An important question is how to input the light pulse into the gas. 
There are different possibilities. One example uses a grid mirror 
that has the grid stripes of small area, so that atoms can freely fly
through the mirror, and small spacing between the grid stripes
as compared to the wavelength of light to provide
efficient reflection, as in Fig.~1b. 
If atoms are at rest, 
the light would propagate in the
forward direction. 
However, if the velocity of 
atoms is equal to (or larger than) $\tilde v_{g}$, one should see 
a frozen (or backward) pulse. 

Depending on the mechanism of pulse input into the medium, one 
should look for the solution of the 
problem with initial (time), boundary (space), 
or mixed (time-space) conditions. 
 In the case of {\em the initial value problem}, 
 we solve the dispersion equation for $\w=\w(k)$, Fig.~3a. 
 Galilean transformation ensures 
 the same EIT half width, $\dk_{EIT} = \W^2/\g\tilde v_g$, 
 as for the atoms at rest since $Im[\tilde\w(\tilde k)]=Im[\w(k)]$.
     In the case of {\em the boundary value problem}, we find $k=k(\w)$. 
The result shows narrowing of the EIT dip 
proportional to the kinematic factor 
$\alpha = (\tilde v_g - v)/\tilde v_g$. 
Indeed, in the accompanying frame the dispersion relation 
near EIT resonance can be decomposed in 
a form of a quadratic polynomial, 
$\tdk = \dk_0- i [\kappa_0 + \xi (\tdw-\dw_d)^2] + 
{(\tdw-\dw_d)/\tilde v_g}$.
Its Galilean transformation to the 
laboratory frame yields 
\be
\dk = \dk_0+
{1\over\alpha}\left[
{\delta\w\over\tilde v_g} -
i\kappa_0 -i\xi\left({\delta\w\over\alpha}\right)^2
\right], 
\ee
where $\delta\w = \dw-\dw_d $. 
Coefficients in Eq.~(3) can be easily deduced using Eq.~(2). For example,
for the case of one-photon resonance $\dw_d=0$ at $\W^2 \gg \g_{cb}\g$,
we have $\tilde v_g=\hbar\W^2/2\pi \mu_{ab}^2k_dN$, 
$\dk_0=0$, 
residual absorption coefficient 
at the center of EIT dip is $\kappa_0 = \g_{cb}/\tilde v_g$, and 
a coefficient determining the parabolic 
profile of absorption in the EIT dip is 
$\xi = \g/\W^2\tilde v_g$.
This approximation is valid if residual absorption is 
small, $\kappa_0\xi \ll (1-v/\tilde v_g)^2/v^2$.
Absorption increases twice as much as EIT minimum value 
at detuning $\delta\w_{EIT}=|\tilde v_g - v|\W\sqrt{\g_{cb}/\g}/\tilde v_g$ that is much less
than the EIT half width 
$\dw_{EIT} = \dk_{EIT}|\tilde v_g - v|$.

Eq.~(3) shows that 
the absorption coefficient $Im\; k$ is increased and sharpened by a factor 
$(\tilde v_g-v)/\tilde v_g$ as compared to that in the co-moving frame. 
Since the spectrum of the pulse cannot be 
transformed on the stationary boundary, only those spectral components
that are within the sharpened EIT dip penetrate deep into the medium. 
For drift velocity $v > \tilde v_g$, 
the backward EIT polariton can be excited from 
inside a cell (Fig.~1b).

     In the case of an atomic beam with a moving boundary (or moving
sample), i.e., for {\em the mixed boundary-initial value problem}, 
the spectrum (inverse duration) of the pulse shrinks at
the moving boundary
exactly in the same way as the EIT width in Eq.~(3), 
$\dw = \tdw{(\tilde v_g - v)/\tilde v_g}$. 
This is not a coincidence, but is necessary 
for consistency of viewing of the same process
from different frames.
The pulse within the EIT dip
decays in time with the same rate 
independently of whether it propagates through atoms at rest or through a beam, 
since this decay is pre-determined by atomic relaxation $\g_{cb}, \g$. 

{\bf Atoms with a thermal velocity distribution.}
Let us consider a stationary cell of hot atoms. 
If the intensity of
the drive is strong enough to provide EIT for the resonant group of atoms
(see Fig.~4)
but at the same time weak enough to avoid an interaction with
off-resonant atoms, moving with ``wrong'' velocities, it is mainly
this drifting beam that 
would support {\it the ultra-slow EIT polariton} with zero or even
negative group velocity.

     To prove this we calculate the dispersion law
$\w(k)$ for the EIT polariton in a hot gas in a cell at rest.
The susceptibility is given by an average of the beam
susceptibility over a velocity distribution $F(v)$
of atoms in a gas with thermal velocity $v_T$,
$
\chi(\w,k) = \int^{+\infty}_{-\infty} dv F(v) \chi_v(\w,k).
$
Instead of the Maxwellian
thermal distribution we can use Lorentzian, 
$F(v) = \ds{v_T/[\pi( v_T^2 + v^2)]}$,
since the far-off-resonant tails are not important. 
This allows us to obtain simple analytical results because an
integration over velocities is reduced to a sum of a few
residues in the simple poles, $v=v_j$. 
 Only those poles count
 that lay in the lower half complex $v$-plane in
 the formal limit of infinitely large growth rate
 $Im \w\rightarrow - \infty$. 
For a positive wavenumber detuning, 
$\dk > 0$, there are two such poles. 
One originates from Lorentzian, 
$v_1 = -iv_T$, 
and the other from the velocity dependent populations, 
$v_2 = -(i\g G + \dw_d)/k_d $. 
Here $\g G = \ds\g\left(1 + \ds{\W^2/(\g_{cb}\g)}\right)^{1/2}$
determines the velocity width of an effective drifting beam of atoms that  
are driven by an external field 
into a coherent ``dark'' state \cite{3,scullybook}, 
and, hence, responsible for the ultra-slow EIT polariton (see Fig.~4). 
For $\dk < 0$, there is 
an additional pole, $v_3 \propto 
1/\dk$, 
originated from resonance $\G_{ab}\G_{cb}+\W^2=0$ in Eq.~(2).
However, near EIT resonance, i.e., for small 
detuning $\dk$, it enters 
the lower half plane from infinity, $v_3 \rightarrow - i\infty$, 
so that its contribution is negligible  if 
$N \ll k_d^3(\g_{cb}/\g)\sqrt{k_dv_T/\W}$.

Calculation of the residues at poles $v_1$ and $v_2$ yields 
\be
\chi(\w, k) 
= {i\mu_{ab}^2 N\over 2\hbar}\left[
{\eta_1
\over
\W^2 + \G^{(1)}_{ab}\G^{(1)}_{cb}}+ 
{\eta_2
\over
\W^2 + \G^{(2)}_{ab}\G^{(2)}_{cb}
}\right],
\ee
where 
$
\eta_1 = [R_1\G_{ac}^{(1)} - 
\G^{(1)}_{cb}( 1+ 
2\g R_1/\g_{cb})
]
/
[ 
1 + \g^2(G^2-1) R_1/\W^2
] 
,
$ 
$
\eta_2 =  k_dv_T R_2 
[
\W^2/(G-1) - 
\G^{(2)}_{cb}\g
]/
\g_{cb} \g G  
$,   
$\G^{(1)}_{ab} = \g + kv_T + i\dw$ ,
$\G^{(2)}_{ab} = \g(1 + Gk/k_d) + i(\dw-{k\dw_d/k_d})$, 
$\G_{ac}^{(1)} = \g + k_dv_T + i\dw_d$,
$\G^{(1)}_{cb} = \g_{cb} + |\dk| v_T + i(\dw-\dw_d)$ ,
$\G^{(2)}_{cb} = \g_{cb} + |\dk|\g G/ k_d + i(\dw-\dw_d-\dk\dw_d/k_d)$,
$R_1 = {\W^2/[\g^2 + (\dw_d - ik_dv_T)^2]}$,
$R_2 = {\W^2/[(k_dv_T)^2 + (\dw_d + i\g G)^2]}$.

The susceptibility (4) 
of a hot gas looks like the susceptibility of a
medium consisting of just 
two mono-velocity components:
(i) broad background with velocity $v=0$ and linewidth $\g + k_d v_T$,  
and (ii) a drifting beam with velocity $v_d = -\dw_d/k_d$  
and power broadened linewidth $\g(1+G)$ (see Fig.~4).
This interpretation  
becomes very accurate near 
EIT dip, $|\dw-\dw_d| \ll \g G$, at the conditions necessary for the existence of 
freezing ultra-slow EIT polariton:
     a) low-frequency coherence decay is much slower than optical
decay ($\g_{cb}\ll\g$);
     b) drifting beam width is less than Doppler broadening
($\g G\ll k_d v_T$);
     c) detuning of driving and probe fields from 
one-photon resonance is large enough ($|\dw_d| \gg \g G$)
while two-photon resonance is maintained.
Then, for the ultra-slow EIT polariton, 
the susceptibility is approximated as 
\be
\chi = 
{\mu_{ab}^2 N'\over \hbar\g G}\left[
{\W^2\over\g(1+G)(\w-\w_{k})} -i\right],
\ee
if we keep only resonant $\w$-dependence 
in denominators setting everywhere else $\dw=\dw_d$.
Here
$N' = N \g G k_dv_T/[ (k_d v_T)^2 + \dw_d^2] \ll N$ is the density of atoms 
in the drifting beam.   
The resonant denominator, where 
$\w_k =\w_{ab} + \dw_d{k/k_d} + i\g_k$, 
$\g_k =\g_{cb} + {\W^2/\g(1+G)} + {|\dk| \g G/ k_d}$,
comes from the factor $\W^2 +\G^{(2)}_{ab}\G^{(2)}_{cb}$ in Eq.~(4). 
Thus, we explicitly find the frequency and the decay ($\w_k$, $\g_k$) of 
{\it the EIT exciton} coupling of which to the probe field 
produces the ultra-slow polariton.

For the boundary value problem, Eq.~(5) yields   
a dispersion that is similar to that for the mono-velocity beam (3) with parameters
$v=v_d$,
$\tilde v_g' = [(k_dv_T)^2 + \dw_d^2]\W^2\hbar/[\mu_{ab}^2N \g(1+G)k_d^2v_T]$,
$\kappa_0 = \g_{cb}/\tilde v_g'$, $\xi = 1/\g_k\tilde v_g'$.

For the initial value problem, from the dispersion equation $kc = \w(1+2\pi\chi)$ and 
Eq.~(5), we find dispersion law
\be
\dw = \dw_d - v_d\dk + i\g_k -
\ds{\W^2\over\g(1+G)}\left[{\hbar\g G\dk\over2\pi\mu_{ab}^2k_dN'}
+i\right]^{-1}
\label{dw}
\ee
shown in Fig.~3b. 
The EIT half width is $\dk_{EIT}'=\g_k/\tilde v_g'$. For small detuning  
$|\dk| \ll \dk_{EIT}'$, 
Eq.~(\ref{dw}) yields linear dispersion and parabolic decay profile,
$\dw=\dw_d+\dk(\tilde v_g'-v_d) + i\g_{cb} + i\dk^2 \tilde {v'}^2/\g_k$. 
Decay increases twice as much as EIT minimum value, $Im \dw 
 = 2\g_{cb}$, at very small detuning 
$\delta k_{EIT}' = \sqrt{\g_{cb}\g_k}/\tilde v_g' \ll \dk_{EIT}'$.   
The group velocity describes pulse kinematics if $d\w/dk$ has 
negligible imaginary part, i.e., near the center of the EIT dip where
$|\dk| < |\tilde v_g - v_d|\g_k/\tilde {v'}^2_g$. 
 The last inequality does not mean that the pulse cannot be stopped.
 It just means that when the pulse is frozen, 
 $v_g=\tilde v_g'-v_d = 0$, its evolution
 is governed by the dispersion of absorption. 

Fig.~3 clearly shows that the ultra-slow 
EIT polariton in a hot gas is similar to that in a mono-velocity beam, since 
detuning of driving field picks a beam 
with velocity $v_d=-\dw_d/k_d$.
However, effective density of atoms supporting EIT polariton  
$N'$ and EIT width 
$\dk'_{EIT} = \g_k/\tilde v_g'$
in a hot gas are different because of factors $\g G$ and $F(v)$. 
As a result, the group velocity 
at the EIT resonance, according to Eq.~(6), in terms of a critical density is  
\be
v_g = {\beta N_{cr}\over NF(v_d)} - v_d, \;\; 
N_{cr} = 
{\hbar\W\over 2\pi^2\beta\mu_{ab}^2}\sqrt{\g_{cb}\over\g},
\ee 
where $\beta=\max[v_dF(v_d)]$. For Lorentzian $F(v_d)$, we have $\beta=1/2\pi$, 
and $v_g =(v_d-v_d^{(1)})(v_d-v_d^{(2)})N_{cr}/2Nv_T$  
is a quadratic polynomial over $v_d$, 
i.e., the group velocity is zero for drive detunings
$v_{d}^{(1,2)} = v_T[N/N_{cr}\pm\sqrt{(N/N_{cr})^2-1}]$ and negative 
between them for density higher than the critical value, $N>N_{cr}$, as is shown in Fig.~2. 
To achieve minimal group velocity,
$\min v_g = -(v_T N/2 N_{cr})[1-(N_{cr}/N)^2]$,
one has to tune at $v_d=v_T N/N_{cr}$. 
The condition to freeze or reverse the light ($v_g \leq 0$) 
means that the group velocity supported by the drifting beam 
with the density $N'=\pi NF(v_d)\g G/k_d$  
should be equal to or less than the velocity of atoms in the beam, 
i.e., $\tilde v_g' = \tilde v_g N'/N \leq v_d$. 
If we compare a mono-velocity beam with a hot gas at $v_d = v$ and the same $N'$ 
as the total density $N$ in a beam to provide the same group velocity, 
$\tilde v_g = \tilde v_g'$, we find that the EIT width and the residual decay in a hot gas 
are $G\simeq \W/\sqrt{\g_{bc}\g}$ times less than in a beam.  
To minimize $N_{cr}$ the drive intensity should be as low as possible 
to decrease $\tilde v_g'$ due to power broadening effect and  
to avoid EIT contribution  
from the atoms with ``wrong'' (positive)  velocities. 
That is, the drive intensity should be just above a threshold 
of the EIT effect at resonance, 
$\W^2 > \g_{cb}\g$. 
Under realistic parameters relevant to the experiments with $^{87}$Rb vapor 
\cite{kash99prl} and chozen in Figs.~2-5, the critical density is 
$N_{cr}\sim 10^{11}$~cm$^{-3}$.

Absorption or time variation of the drive field results in 
a spatial or time dependence 
of the group velocity in the cell. 
This allows us to control input and parameters of the pulse in the cell. 
According to geometrical optics,
the parameters of the EIT polariton adiabatically
follow the local properties of the driven atoms. 
Fig.~5 demonstrates how the
ultra-slow pulse decelerates up to the point $v_g=0$ where it becomes frozen. 

The important conclusion is that 
the drifting beam provides large enough drift spatial dispersion 
$\dd n/\dd k$ (see Eq.(1)) to ensure $v_g \le 0$.
Although the density of drifting atoms is small 
$N' \ll N$, their resonant contribution dominates. 
This allows us to make the group velocity
zero or even negative \cite{lastnote}.  
To observe freezing or backward light
one can look, e.g., for a scattering, 
luminescence, delay, 
or enhanced nonlinear mixing caused
by ultra-slow pulse.

We thank M.~Fleishhauer, E.\ Fry, S.\ Harris, 
M.~Lukin, 
and G.\ Welch for 
helpful discussions.
This work was supported by the ONR,
the NSF, the Welch Foundation,  
and the Texas Advanced Technology Program.


\def\etal{\textit{et al.}}


\begin{figure}[ht]
 \centerline{
} \vspace*{2ex}
\label{fig1}
 \caption{
 	\label{setups.fig}
(a) Three-level atomic $\Lambda$-system.
(b) Geometry of ultra-slow EIT pulse propagation in the gas of atoms. 
}
\end{figure}


\begin{figure}[ht]
 \centerline{
}
 \vspace*{2ex}
 \caption{
 	\label{fig2}
Ultra-slow and negative
group velocity of EIT polariton vs detuning of drive laser;
$\W=0.25 \g$, 
$k_dv_T=100\g$, $\g_{cb}=0.001\g$, 
(a) $N =0.6N_{cr}$;
(b) $N =N_{cr}$;
(c) $N =1.5N_{cr}$.
%
}
\end{figure}


\begin{figure}[ht]
\center{
}
 \vspace*{2ex}
 \caption{
 	\label{fig3}
Dispersion, $Re\dw=Re[\w-\w_{ab}]$,  
and decay, $Im\dw$, spectra of the ultra-slow EIT polariton 
according to numerical solution of the dispersion equation
for: (a) atomic beam 
($N = 1.1 N_{cr}\pi F(v_d)\g G/k_d$)
with susceptibility~(2); 
(b) stationary cell of hot gas ($N=1.1N_{cr}$)
with exact susceptibility (4); 
$\W=0.25 \g$, 
$v=v_d=v_T$, 
$k_dv_T=100\g$, 
$\g_{cb}=0.001\g$. 
%
%
}
\end{figure}


\begin{figure}[ht]
\center{
}
 \vspace*{2ex}
 \caption{
 	\label{fig4}
The velocity distribution of atoms in a cell (solid line). 
Effective drifting beam (dotted) selected by 
drive laser.
}
\end{figure}


\begin{figure}[ht]
\center{
}
 \vspace*{2ex}
 \caption{
 	\label{fig5}
Kinematics of the 
deceleration of the ultra-slow pulse 
to the point of freezing ($v_g=0$) along a cell with 
decreasing group velocity $v_g(z)$. 
Positions of pulse are shown at subsequent moments of time $t=m\tau$
($\tau=3L/2v_g(0)$, m=0,1,2,3). $v_g(z)$ is calculated numerically 
according to decreasing drive intensity found from the wave equation
for the same parameters as in Fig.~3(b), $L=10$ cm.
%
}
\end{figure}



\end{document}